\documentclass[12pt]{article}
\usepackage[dvips]{graphicx}
\usepackage{latexsym}
\usepackage{amssymb}

\newcommand{\R}{\mathbb{R}}

\def\p{\partial}

\newcommand{\bea}{\begin{eqnarray}}
\newcommand{\eea}{\end{eqnarray}}

\newcommand{\koniec}{\begin{flushright}  $\Box $ \end{flushright}}
\def\be{\begin{equation}}
\def\ee{\end{equation}}

\def\const{\mbox{const}}

\def\p{\partial}

\newtheorem{theo}{Theorem}[section] 
\newtheorem{prop}[theo]{Proposition}  

\newtheorem{defi}[theo]{Definition}

\newcounter{mnotecount}[section]

\renewcommand{\themnotecount}{\thesection.\arabic{mnotecount}}

\newcommand{\mnote}[1]
{\protect{\stepcounter{mnotecount}}$^{\mbox{\footnotesize
$
\bullet$\themnotecount}}$ \marginpar{
\raggedright\tiny\em
$\!\!\!\!\!\!\,\bullet$\themnotecount: #1} }

\begin{document}
\title{\vskip -70pt
\begin{flushright}
{\normalsize DAMTP-2011-23} \\
\end{flushright}
\vskip 60pt
{\bf  Optical Metrics and Projective Equivalence \vskip 15pt}}
\author{Stephen Casey\thanks{Email: sc581@cam.ac.uk}, 
Maciej Dunajski\thanks{Email: M.Dunajski@damtp.cam.ac.uk},
Gary Gibbons\thanks{Email: G.W.Gibbons@damtp.cam.ac.uk},
Claude Warnick\thanks{Email: cmw50@cam.ac.uk}\\
Department of Applied Mathematics and Theoretical Physics,\\
University of Cambridge,
Wilberforce Road, Cambridge CB3 0WA, UK. 
}
\date{}
\maketitle
\begin{abstract}

Trajectories of light rays in a static spacetime are described by
unparametrised geodesics of the Riemannian optical metric associated with the Lorentzian  spacetime metric. We investigate 
the uniqueness of this structure and demonstrate that two different observers, moving relative to one another, who both see the universe as static may determine the geometry of the light rays differently. More specifically, we classify Lorentzian metrics admitting more
than one hyper--surface orthogonal time--like Killing vector and
analyze the projective equivalence of the resulting optical metrics.
These metrics are shown to be projectively equivalent up to diffeomorphism if the static Killing vectors generate a group $SL(2, \R)$, but not projectively equivalent in general. We also  consider the cosmological $C$--metrics in Einstein--Maxwell theory and 
demonstrate that optical metrics corresponding to different values
of the cosmological constant are projectively equivalent. 
\end{abstract}
\section{Introduction--nonequivalent optical metrics}
When trying to interpret the physical properties of a spacetime,
of fundamental  importance is the behaviour of null geodesics as these correspond to the  trajectories of light rays. The vast majority of measurements made of the universe consist of observation of electromagnetic waves emitted in the past at great distances from us. The behaviour of light rays as they bend around the  sun gave the first observational evidence for General Relativity and such gravitational lensing continues to be a significant branch of
astronomy.

In the case where a spacetime is static or conformally static, a powerful approach for investigating the properties of light rays is the \emph{optical metric}. This may be thought of as a natural Riemannian geometry experienced by light rays. It has been recently used to study light bending by a black hole in the presence of a cosmological constant \cite{GWW08} and to give an alternative interpretation of black hole no-hair theorems \cite{GW}. An important question one should address when introducing such a structure is to what extent it is unique, in other words can
one spacetime give rise to more than one optical metric. Physically this would mean that there exist two different observers, moving relative to one another, who both see the universe as (possibly conformally) static and who would determine the geometry of the light rays differently. This is the question we shall address here.

 Let $(M, g)$ be a pseudo--Riemannian manifold with a metric
of signature $(D, 1)$, where $D>0$. The metric is called static if it admits
a hyper--surface--orthogonal (HSO) time--like Killing vector $K$. Any such metric
is locally of the form
\be
\label{metric}
g=V^2(-dt^2+h),
\ee
where $h=h_{ij}dx^idx^j$ is a Riemannian metric on the space of orbits $\Sigma$
of $K=\p/\p t$ and $V=V(x^i)$ is a function on $\Sigma$.
The metric $h$ is called the optical metric of $g$
and the motivation behind this terminology \cite{abramowicz,GWW08,GW} comes 
from the fact that 
null  geodesics of $g$ project to unparamertrised  geodesics of $h$. This can be readily verified
as null geodesics of $g$ coincide with the null geodesics of $V^{-2}g$. 

\begin{figure}
\begin{center}
\caption{Non equivalent optical metrics}
\vskip10pt
\label{fig1}
\includegraphics[width=10cm,height=5cm,angle=0]{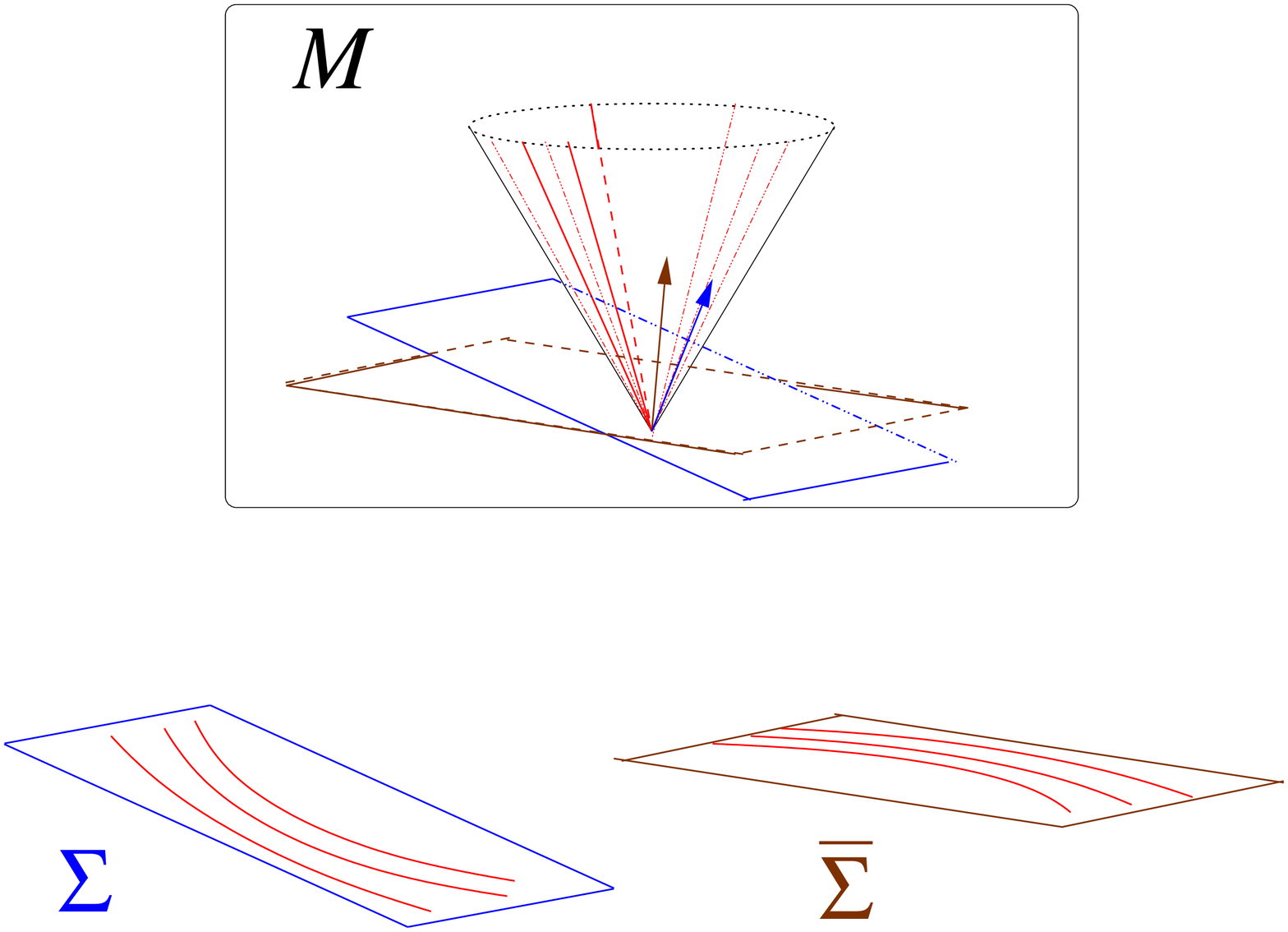}
\end{center}
\end{figure}

 It is clear from this discussion that an optical metric depends on the choice
of a static time--like Killing vector (Figure \ref{fig1}).  
Three different equivalence classes of Riemannian metrics will play a role 
in our discussion. Let $(\Sigma, h)$ and $(\bar{\Sigma}, \bar{h})$ be
two $D$--dimensional Riemannian manifolds, and let $\rho:\Sigma\rightarrow
\bar{\Sigma}$ be a diffeomorphism. The metrics $h$ and $\bar{h}$ are 
\begin{itemize}
\item {\it Equivalent}, if there exists a $\rho$ such that
$\rho^* \bar{h}=h$.
\item {\it Projectively equivalent}, if there exists a $\rho$
such that $\rho^* \bar{h}$ and $h$ share the same unparametrised 
geodesics.
\item{\it Optically equivalent}, if there exists a pseudo--Riemannian
$(D+1)$--dimensional manifold $M$ with two HSO Killing vectors $K$ and
$\bar{K}$ such that $\Sigma$ and $\bar{\Sigma}$ are 
hyper--surfaces orthogonal to $K$ and $\bar{K}$ respectively, and
$(h, \bar{h})$ are optical metrics of  $K$ and $\bar{K}$ respectively.
\end{itemize}
All equivalences we shall discuss are in fact local equivalences as $\rho$ is only required to be a smooth map between some open sets.
 
  If two metrics are equivalent, they are also projectively equivalent,
but the converse is not true in general. In this paper we shall
analyse the connection between the projective equivalence and optical 
equivalence. It turns out that the latter 
{\it almost always} implies the former.

Let us assume that $(M, g)$ admits two optical metrics 
 $h$ and $\bar{h}$. Thus $g$ can be written in the form (\ref{metric})
in more than one way. Therefore there exists
a diffeomorphism $f:M\rightarrow M$ such that $f^*g$ and $g$ are both of the form
(\ref{metric}) albeit written in different coordinate systems
\[
V^2(-dt^2+h)=\bar{V}^2(-d\bar{t}^2+\bar{h}),
\]
where $\bar{V}=\bar{V}(\bar{x})$ and $\bar{x}=\bar{x}(x, t), \bar{t}=\bar{t}(x, t)$.
Moreover  $\bar{K}=\p/\p \bar{t}$ and $K=f_*(\p/\p t)$ are two time--like
HSO Killing vectors.
If one of these vectors is a constant multiple of the other 
then we can deduce that the optical metrics $h$ and $\bar{h}$ are related by a 
constant rescaling. Let us therefore assume that these vectors
are not proportional.

  We emphasise that the light cone structure on $M$
does not give rise to a canonical 
bijection between geodesics of $h$ and $\bar{h}$.
\begin{figure}
\begin{center}
\caption{No bijection between geodesics of two optical metrics}
\vskip10pt
\label{fig2}
\includegraphics[width=10cm,height=6cm,angle=0]{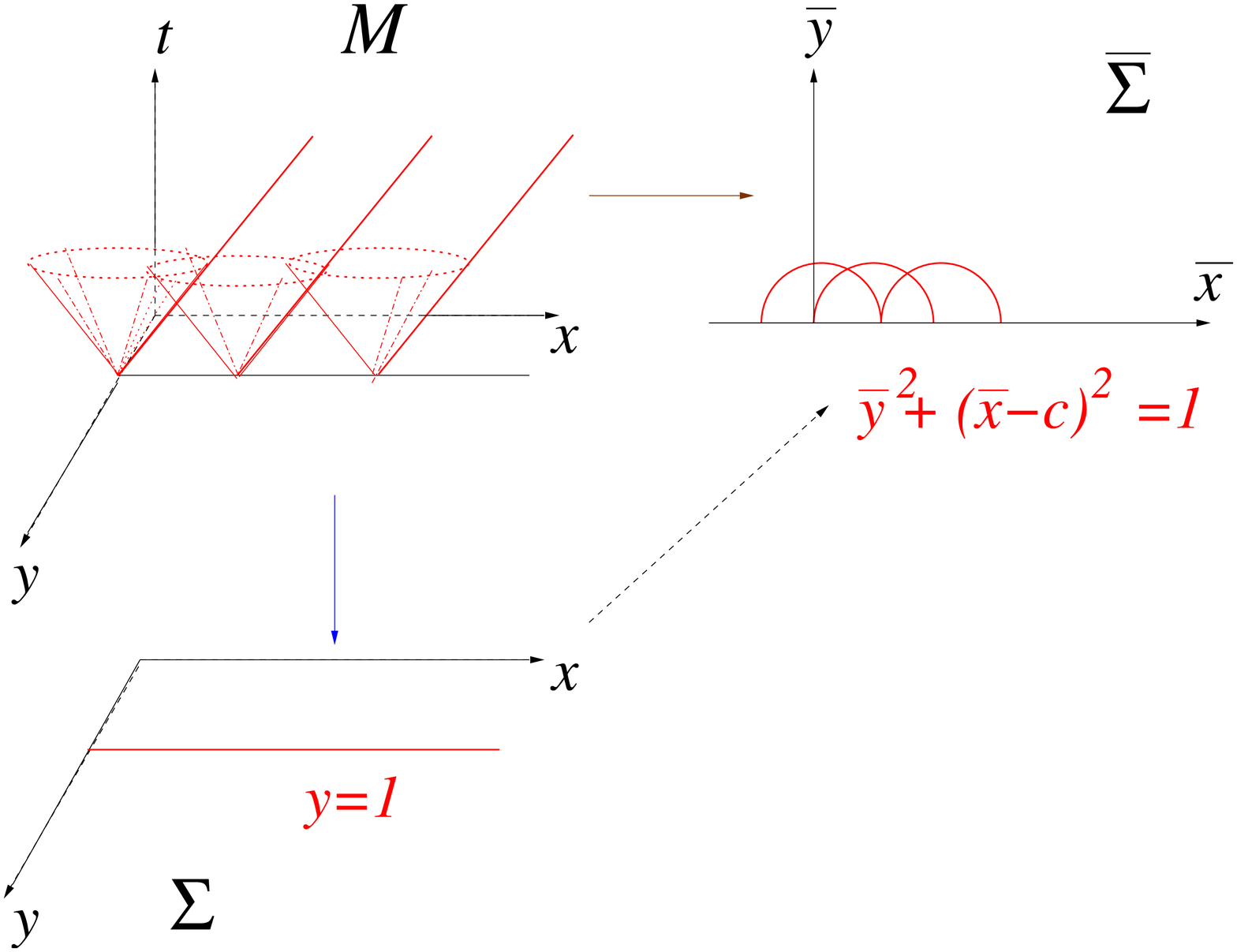}
\end{center}
\end{figure}
For example, if  
\[
g=-dt^2+dx^2+dy^2
\]  
is the Minkowski metric
on $M=\R^{2,1}$, and $K=\p/\p t$ then the associated optical metric
is $h=dx^2+dy^2$. 
Setting 
\[
t=\bar{y}\sinh{\bar{t}},\quad x=\bar{x},\quad  y=\bar{y}\cosh{\bar{t}}
\]
yields $g=\bar{y}^2(-d\bar{t}^2+\bar{h})$, where
the upper half-plane metric $\bar{h}=\bar{y}^{-2}(d\bar{x}^2+d\bar{y}^2)$
is the optical metric of $\bar{K}=\p/\p\bar{t}$. Now consider a geodesic
$\gamma$ of $h$ given by $y=1$. This lifts to a one parameter family
of null geodesics $\{y=1, t=x-c\}$ of $g$, and this family projects
to a family $\gamma_c$ of geodesics of $\bar{h}$ given by unit semicircles
\[
\bar{y}^2+(\bar{x}-c)^2=1
\]
parametrised by the position of their centres on the $\bar{x}$ 
axis (Figure \ref{fig2}) .

Note that for this example $h$ and $\bar{h}$ are projectively equivalent:
there exists a diffeomorphism between the Euclidean plane $\Sigma=\R^2$
and the upper half plane $\bar{\Sigma}=\mathbb{H}^2$ which maps unparametrised
geodesics of $h$ to unparametrised geodesics of $\bar{h}$.
We  shall demonstrate that this is not the case in 
general.

 The paper is organised as follows: 
In Section \ref{secmulti} (Proposition \ref{HSOone})  we  shall find generic local forms of Lorentzian metrics
which admit two non proportional HSO timelike Killing 
vectors\footnote{This problem was already 
addressed in \cite{S10} but our 
construction will be different. In the language of  \cite{S10} 
we shall look for a special case of conformal ultrastatic transformations such
that $\exp{2\Phi}=\bar{V}^2/V^2$.}. They are warped product metrics
on $M=S_0\times S_1$ given by
\be
\label{metric_final}
g= e^w\gamma_0+\gamma_1
\ee
where $(S_0, \gamma_0)$ is a two--dimensional Lorentzian manifold  
of constant curvature,  $(S_1, \gamma_1)$ is an arbitrary 
two--dimensional Riemannian manifold and $w:S_1\rightarrow \R$ is an 
arbitrary function.  We shall also show that imposing the Einstein condition
on (\ref{metric_final}) leads to non--trivial metrics which are analytic
continuations of the Kottler metric (Proposition \ref{prop_einstein}).
In Section \ref{secoptical} we shall  compute the optical metrics associated 
to  each  Killing vector (Proposition \ref{proj} and Proposition  \ref{reproj}).
In Section \ref{secprojective} we shall determine when optically equivalent 
metrics are projectively equivalent. If the curvature of $\gamma_0$ is non--zero, then the general HSO Killing time--like vector is a linear combination of the generators of $SL(2, \R)$ acting  isometrically on $M$ with two--dimensional orbits $S_0$. In this case the resulting optical metrics are projectively equivalent to
\[
h=(1 -\kappa r^2)^{-2}dr^2+e^{-w}(1-\kappa r^2)^{-1}\gamma_1,
\] 
where  $\kappa=\pm 1$ is the curvature of $\gamma_0$ (Proposition \ref{quiv}). If $\gamma_0$ is flat, then the HSO Killing vector arises from the generators of the three--dimensional  group $Sol$ of isometries of $\R^{1, 1}$ and the optical metrics are not projectively equivalent in general.
In Section \ref{seccmetric} we shall consider the cosmological $C$--metrics in Einstein--Maxwell theory. These metrics fall outside of our
class (\ref{metric_final}), and the notion of optical metric is unambiguous. We shall demonstrate that optical metrics corresponding to different values
of the cosmological constant are projectively equivalent. 
Thus the trajectories of light--rays in the $C$--metric space--times
depend on the mass and electric charge, but not on the cosmological constant.
\vskip5pt{\bf Acknowledgements.} 
MD is grateful to Vladimir Matveev for helpful correspondence, and to IPM in Tehran
where some of this work was done for hospitality. 
\section{Multi--static metrics}
\label{secmulti}
We shall now classify local forms of pseudo--Riemannian structures 
$(M, g)$ which
admit more than one HSO time--like Killing vector.
\begin{defi}
A Lorentzian metric is called multi--static if it admits at least two 
non--proportional
HSO time--like Killing vectors.
\end{defi}
From now on we shall assume that the dimension of $M$
is equal to four.
 Let $(K, \xi)$ be two HSO time like Killing vectors\footnote{The vector field 
 $\xi$ is closely related to $\bar{K}$ from the previous section. These vector fields are however defined on different spaces which justifies our notation. Moreover, 
 as we shall see in Section \ref{secoptical}, the general
 form of $\xi$ depends on some constants of integration and thus
several different forms of $\bar{K}$ can arise.}  on $M$. We can choose 
a local coordinate system 
(\textbf{Note:} We use  the  letters from the start of 
the alphabet ($a,b,c,\ldots$) to run over 0,1,2,3 and letters from the middle of the alphabet ($i,j,k,\ldots$) to run over 1,2,3)
$x^a= (t, x^i), $ such that the metric  
is given by (\ref{metric}) and  $K=\p/\p t$. In this coordinate system
\[
\xi=\xi^0\frac{\p}{\p t}+\xi^i\frac{\p}{\p x^i},
\]
where $\xi^0, \dots, \xi^3$ are functions of $(x, t)$. From our assumptions 
it follows that not all $\xi^i$ are identically zero 
(if they where, then the Killing equations 
$\nabla_0\xi_0=\nabla_{(i}\xi_{0)}=0$ would imply $\xi^0=\const$ thus contradicting our assumptions about the independence of $K$ and $\xi$). Therefore
there exists $t_0$ such that the projection of the restriction of $\xi$ at
the surface $\Sigma$ given by $t=t_0$
\be
\label{new_eq}
\tilde{\xi}=\xi|_{t=t_0}
\ee
is a non--zero vector field.  Furthermore, we can make the coordinate transformation $t \rightarrow t - t_0$ while preserving the form of the metric (\ref{metric}) so that $\tilde{\xi}^i = \xi^i\mid_{t=0}$.

The HSO Killing equations for $\xi$  imply that
$\tilde{\xi}$ is a HSO Killing vector for $V^2h$ and so there exists a function
$r:\Sigma\rightarrow \R$ such that
\[
V^2h=e^w dr^2+\gamma,
\]
where $\tilde{\xi}=\p/\p r$, and $(w, \gamma)$ are a function and a metric
on a two--dimensional surface $S_1$ (the space of orbits of 
$\tilde{\xi}$ in $\Sigma$) which do not depend on $r$. We can
use the isothermal coordinates $(x, y)$ so that $\gamma=e^u(dx^2+dy^2)$
and $u, w$ are functions of $(x, y)$. Thus the most general Lorentzian 
metric which admits more than one optical metric is locally of the form
\be
\label{canonical}
g=-V^{2}dt^2+e^wdr^2+e^u(dx^2+dy^2),
\ee
where $V=V(r, x, y)$,  $u=u(x, y)$ and $w=w(x, y)$. We note that
the function $V$ is not arbitrary - its form is restricted by the 
Killing equation for $\xi$. 

Our next step is to classify the normal forms of $\xi$ and thus
read off the canonical forms of its optical metric 
$\bar{h}$ on some three--manifold $\bar{\Sigma}$ where $\bar{K}=\p/\p\bar{t}$
giving rise to $\bar{h}$ is the push forward of  $\xi$ under some local diffeomorphism between $\Sigma$ and $\bar{\Sigma}$.
We shall  make the 
additional genericity assumption 
\begin{defi}
A multi--static metric is called generic if the isometry group 
generated by any pair  of HSO time--like Killing vectors (and their commutators)
has two--dimensional orbits in $M$.
\end{defi}
The genericity assumption implies that for any $t_0$, the HSO Killing vector $\xi$ restricted to the surface $t=t_0$ 
defined by $K$ is proportional to a fixed vector field.
\begin{prop}
\label{HSOone}
Any generic multi--static metric is locally a warped product metric on $M = S_0 \times S_1$ given by
\be
\label{prof_metric}
g = e^w \gamma_0 + \gamma_1
\ee
where $(S_0, \gamma_0)$ is a two-dimensional Lorentzian manifold whose curvature is constant, $(S_1, \gamma_1)$ is a two-dimensional Riemannian manifold  and $w:S_1\rightarrow \R$ is an arbitrary function.
\end{prop}
{\bf Proof.}
First we shall show that given a pair of HSO time--like Killing vectors 
$(K, \xi)$, the genericity assumption implies
existence of two functions $(r, t)$ such that the metric 
takes the form  (\ref{canonical}), 
and
\begin{equation}
K=\frac{\p}{\p t}, \quad \xi = \xi^0(t,r,x,y) \frac{\partial}{\partial t} 
+ a(t) \frac{\partial}{\partial r}
\label{formone}
\end{equation}
where $(x, y)$ are coordinates on surface $S_1$ parametrising the 2D orbits in $M$, and $a$ is a function which depends only on $t$.
To prove this statement, note that the group generated by
the Killing vectors and their commutators acts on $M$ 
with two--dimensional orbits so
\be
\label{condi_0}
[K, \xi]=pK+q\xi,
\ee
where $p, q$ are functions on $M$. We need to show that there exists
functions $\alpha, \beta$ such that
\be
\label{condi_1}
[\beta^{-1}(\xi-\alpha K), K]=0,
\ee
as then the local existence of $r, t$ will follow from the Frobenius theorem.
Expanding the Lie bracket (\ref{condi_1}) and  using (\ref{condi_0})
gives a pair of ODEs
\[
K(\beta^{-1})=\beta^{-1}q, \quad K(\alpha\beta^{-1})=\beta^{-1}p.
\]
The existence of $\alpha, \beta$ is a consequence of the Picard existence 
theorem applied to these ODEs and
\[
K=\frac{\p}{\p t}, \quad \xi=\alpha K+\beta\frac{\p}{\p r}.
\] 
Now consider the HSO Killing vector $\tilde{\xi}$ given by 
(\ref{new_eq}) on the surface $\Sigma$ of constant $t$. The Killing equations on $\Sigma$ imply that $\beta=\beta(r, t)$ and that for any value of $t_0$ the resulting vector is proportional to the same Killing vector. Thus
$\beta(r, t)= a(t)b(r)$. We now redefine the $r$ coordinate to set $b(r)=1$. This establishes (\ref{formone}). 
 Therefore, for any value of $t_0$,
\[
\xi^i \frac{\partial}{\partial x^i}\mid_{t=t_0} \propto  \frac{\partial}{\partial r}.
\]
The Killing equations $\nabla_{(2} \xi_{0)} = 0 =  \nabla_{(3} \xi_{0)}$
give $\xi^0=\xi^0(t, r)$.
Using this and equation (\ref{formone}) above, the hypersurface orthogonality conditions
$\xi_{[0} \nabla_1 \xi_{2]} = 0$ and 
$\xi_{[0} \nabla_1 \xi_{3]} = 0$ yield
\begin{equation}
V^2(r,x,y) = v^2(r) e^{w(x,y)}
\label{break}
\end{equation}
for some function $v(r)$. Hence, the metric $g$ may already be written as
(\ref{prof_metric}) where the two-dimensional metric $\gamma_0$ 
is given by
\[
\gamma_0 = -v^2 (r) dt^2 + dr^2.
\]
The scalar curvature of this metric is
\begin{equation}
\kappa = -\frac{2 v''(r)}{v(r)}.
\label{scalcurv}
\end{equation}
This will be important later. The only remaining equations that need to be satisfied are the Killing conditions
$\nabla_{(0} \xi_{0)} = 0 $ and $\nabla_{(1} \xi_{0)} = 0$.
These equations give
\[
- v^2(r) \partial_t \xi^0 = v(r) \frac{d v(r)}{dr} a(t),\quad
- v^2(r) \partial_r \xi^0 =
 - \frac{da(t)}{dt}. 
\]
Differentiating the first condition with respect to $r$ and the second condition 
with respect to $t$ and equating the mixed partial derivatives of $\xi^0$ yields
\begin{equation}
\frac{1}{a(t)}\frac{d^2 a(t)}{dt^2} =  \left(\frac{dv(r)}{dr} \right)^2 - v(r) \frac{d^2 v(r)}{dr^2}.
\label{atter}
\end{equation}
The left hand side of this equation is a function of $t$ only. Hence
\[
\Big(\frac{dv(r)}{dr} \Big)^2 - v(r) \frac{d^2 v(r)}{dr^2} = \Omega = \mbox{constant}.
\]
Differentiating with respect to $r$, we find that
\[
0 = v'(r)v''(r) - v(r)v'''(r) = \frac{v^2(r)}{2} \frac{\partial}{\partial r} \left( - \frac{2v''(r)}{v(r)} \right).
\]
Hence, by (\ref{scalcurv}), the curvature of $\gamma_1$ is constant. Furthermore, if the curvature is $\kappa \neq 0$ then we can set its absolute value to one
by adding a constant to the function $w$. \koniec

\subsection{Einstein Equations}
We shall now impose the Einstein condition
on (\ref{metric_final}) and show that the resulting metrics are analytic continuations of the cosmological Kottler solution.
\begin{prop}
\label{prop_einstein}
Let ${\gamma_0}^{(0)}$ be two dimensional Minkowski metric and ${\gamma_0}^{(\pm 1)}$ be the line element of the two dimensional de Sitter and anti-de Sitter metrics with cosmological constant $\pm 1$  respectively. Consider metrics of the form
\begin{equation}
g = \gamma_1+ w^2 {\gamma_0}^{(k)}
\end{equation}
where $\gamma_1$ and $w$ are respectively a metric and non-constant function on some two dimensional surface. Any such metric which is Einstein, with cosmological constant $\Lambda$ is locally diffeomorphic to the metric
\begin{equation}
\label{claude_metric}
g= \left( k + \frac{c}{r} - \frac{\Lambda}{3} r^2 \right) d\tau ^2 + \frac{dr^2}{ k + \frac{c}{r} - \frac{\Lambda}{3} r^2 } + r^2 {\gamma_0}^{(k)}
\end{equation}
for some constant $c$. The case where $w$ is constant yields that $\gamma_1$ is  an Einstein metric with appropriate cosmological constant to match that of the other factor. 
\end{prop}
{\bf Proof.} The derivation of (\ref{claude_metric}) is analogous to the proof
of Birkhoff's theorem in General Relativity (see, e.g. \cite{HE}),  
except that
the constant curvature warped factor is Lorentzian rather than Riemannian.
One chooses a coordinate system $(r, \tau)$ on $S_1$, where
$r=w$, establishes the $\tau$--independence of the metric and finally examines
the $(r \tau)$ component of the Einstein tensor, which gives the $r$--dependence. 
\koniec
\section{Optical metrics}
\label{secoptical}
To determine the optical metrics resulting from (\ref{metric_final})
we need to consider three cases depending on the curvature of $\gamma_0$.
\subsubsection*{Zero Curvature Case}
We can find local coordinates such that $\gamma_0=-dt^2+dr^2$, and the general HSO Killing vector of $g$ becomes
\[
\xi= (A r + B) \frac{\partial}{\partial t} + (A t+ C) \frac{\partial}{\partial r}
\]
for some constants $A$, $B$ and $C$. 
If $A\neq 0$ we translate $(r, t)$ by adding constants and rescale the Killing vector so that 
\[
\xi=r\frac{\p}{\p t}+t\frac{\p}{\p r}.
\]
Setting $t=\bar{r}\sinh{(\bar{t})}, r=\bar{r}\cosh{(\bar{t})}$ gives the optical metric
of $\p/\p \bar{t}$
\be
\label{opt_ko1}
\bar{h}=\bar{r}^{-2}(d\bar{r}^{2}+e^{-w}\gamma_1).
\ee
If $A=0$ then a constant rescaling of $t$ can be used to set 
$\xi=\cos{\theta}\p_t+\sin{\theta}\p_r$, where $\theta$ is a constant
in a range which makes  $\xi$ is time--like. The pseudo--orthogonal transformation 
of $(r, t)$ can now be used to set $\xi=\p/\p t$, so the optical metric in this case is
\be
\label{opt_ko2}
h=dr^2+e^{-w}\gamma_1.
\ee
\subsubsection*{Anti de Sitter Case}
Now, let us consider the case where the metric has the form (\ref{metric_final}),
where the  constant
curvature of $\gamma_0$ is negative. 
In the $AdS_2$ case we can choose local coordinates so that
\[
\gamma_0 = \frac{-dt^2 + dr^2}{r^2}.
\]
Both $\gamma_0$ and the resulting Lorentzian metric $g$ have three
Killing vectors generating $SL(2, \R)$. In the chosen coordinates these
vectors are
\[
K_1 = \frac{\partial}{\partial t} \,\,\,\,,\,\,\,\, K_2 = t\frac{\partial}{\partial t} + r \frac{\partial}{\partial r} \,\,\,\,,\,\,\,\, K_3 = \left( \frac{t^2+ r^2}{2} \right) \frac{\partial}{\partial t} + t r \frac{\partial}{\partial r},
\]
and
\[
[K_1, K_2]=K_1, \quad [K_2, K_3]=K_3, \quad [K_1, K_3]=K_2.
\]
Furthermore, it is easy to show that any linear combination
\[
\xi = A K_1 + B K_2 + C K_3
\]
is an HSO Killing vector for the metric $g$, which is time--like in some open 
set to which we restrict our attention from now on.
\begin{prop}
\label{proj}
For any timelike HSO Killing vector, $\xi$, of the metric 
{\em (\ref{metric_final})}, where $\gamma_0$ has negative constant 
curvature, the optical metric associated to $\xi$ is diffeomorphic to
\begin{equation}
\bar{h} = \frac{1}{(\phi + \bar{r}^2)^2} d\bar{r}^2 + 
\frac{e^{-w}}{\phi + \bar{r}^2} \gamma_1
\label{diffopt}
\end{equation}
for some constant $\phi$.
\end{prop}
{\bf Proof.} Let us first consider the HSO Killing vectors for 
which $C \neq 0$. Then, adding a constant to $t$ we can set $B=0$ without 
changing the metric. If $A=0$ then divide $\xi$ by $C/2$ to set $C=2$.
Otherwise, rescale $(t, r)$ by the same constant factor to set $A=\pm C/2$
and then divide $\xi$ by $C/2$. Thus the resulting Killing vector 
can take one of three possible forms
\[
\xi = \left(c + t^2 + r^2 \right) \frac{\partial}{\partial t} + 2tr \frac{\partial}{\partial r}, \quad\mbox{where}\quad c=0,-1, 1.
\]
We look for a coordinate transformation $(t, r)\rightarrow (\bar{t}, \bar{r})$
such that $\xi=\p/\p \bar{t}$.
\begin{itemize}
\item If $c=1$ set 
\[
t = \frac{\sqrt{\bar{r}^2 + 4} \cos (2\bar{t})}{\bar{r} - \sqrt{\bar{r}^2+4}
 \sin (2\bar{t})}, \quad
r= \frac{2}{\bar{r} - \sqrt{\bar{r}^2+4}\sin (2\bar{t})}.
\]
\item If $c=-1$ set 
\[
t= \frac{\sqrt{\bar{r}^2-4}(1-e^{4\bar{t}})}{\sqrt{\bar{r}^2-4}(1+e^{4\bar{t}})
-2\bar{r}e^{2\bar{t}}},\quad
r= \frac{4e^{2\bar{t}}}{\sqrt{\bar{r}^2-4}(1+e^{4\bar{t}})-2\bar{r}e^{2\bar{t}}}.
\]
\item If $c=0$ set
\[
t = \frac{\bar{r}^2 \bar{t}}{1-\bar{r}^2 \bar{t}^2},\quad
r = \frac{\bar{r}}{\bar{r}^2 \bar{t}^2 -1}.
\]
\end{itemize}
This gives, in all three cases  
$\gamma_0=-(\bar{r}^2+4c)d\bar{t}^2+(\bar{r}^2+4c)^{-1}d\bar{r}^2$
and the optical metric (\ref{diffopt}) with $\phi=4c$.

 Now consider the case $C=0$. Adding an appropriate constant to $t$ sets
$B=0$ so that
\[
\xi=t\frac{\p}{\p t}+r\frac{\p}{\p r}.
\]
Setting
\[
t=\frac{\bar{r}}{\sqrt{\bar{r}^2-1}}e^{\bar{t}},\quad
r=\frac{1}{\sqrt{\bar{r}^2-1}}e^{\bar{t}}
\]
yields $\xi=\p/\p\bar{t}$ and 
$\gamma_0=-(\bar{r}^2-1)d\bar{t}^2+(\bar{r}^2-1)^{-1}d\bar{r}^2$.
The optical metric in this case is (\ref{diffopt}) with $\phi=-1$.

 Finally, suppose $C = B = 0$ so that 
$\xi = \frac{\partial}{\partial t}$. This gives the optical metric 
\[
\bar{h}=dr^2 + r^2 e^{-w(x,y)} \gamma_1.
\]
A coordinate transformation $r =\bar{r}^{-1}$ 
puts it in the form (\ref{diffopt}) with $\phi = 0$. 
Thus, we have covered all cases.
\koniec
\subsubsection*{de Sitter Case}
In this case $\gamma_0$ can be written in local coordinates as
\[
\gamma_0 = \frac{-dt^2+dr^2}{t^2}.
\]
This switches the role of $r$ and $t$ in the previous section. 
The general HSO timelike Killing vector on $g$ is of the form
\[
\xi = A K_1 + B K_2 + C K_3,
\]
where 
\[
K_1 = \frac{\partial}{\partial r} \,\,\,\,,\,\,\,\, 
K_2 = r\frac{\partial}{\partial r} + t \frac{\partial}{\partial t} \,\,\,\,,\,\,\,\, K_3 = \left( \frac{t^2+ r^2}{2} \right) \frac{\partial}{\partial r} 
+ t r \frac{\partial}{\partial t}.
\]
If $C\neq 0$, then adding a constant to $r$ can be used to set $B=0$. 
The resulting vector will be time--like (in a certain open set in $M$) only
if $AC<0$. In this case we can rescale $(r, t)$ by  the same constant factor
to set $A=-C/2$, so that
\[
\xi = \left(-1 + t^2 + r^2 \right) \frac{\partial}{\partial r} + 
2tr \frac{\partial}{\partial t}.
\]
A coordinate transformation
\[
t= \frac{\sqrt{4-\bar{r}^2}(1+e^{4\bar{t}})}{\sqrt{4-\bar{r}^2}(1-e^{4\bar{t}})
+2\bar{r}e^{2\bar{t}}},\quad
r= -\frac{4e^{2\bar{t}}}{\sqrt{4-\bar{r}^2}(1-e^{4\bar{t}})+2\bar{r}e^{2\bar{t}}}
\]
gives $\xi=\p/\p\bar{t}$ and
\[
\gamma_0=-(4-\bar{r}^2)d\bar{t}^2+\frac{1}{4-\bar{r}^2}d\bar{r}^2
\]
which is defined for  $|\bar{r}|<2$. The optical metric is 
\[
\bar{h}=\frac{1}{(4-\bar{r}^2)^2} d\bar{r}^2 +   
\frac{e^{-w}}{4-\bar{r}^2} \gamma_1. 
\]

If $C=0$ then, adding an appropriate constant to $r$ gives $\xi=K_2$. 
The transformation
\[
t=\frac{\bar{r}}{\sqrt{1-\bar{r}^2}}e^{\bar{t}},\quad
r=\frac{1}{\sqrt{1-\bar{r}^2}}e^{\bar{t}}
\]
yields $\xi=\p/\p\bar{t}$ and 
$\gamma_0=-(1-\bar{r}^2)d\bar{t}^2+(1-\bar{r}^2)^{-1}d\bar{r}^2$.
The optical metric is this case is
\be
\label{opt_c0}
\bar{h}=\frac{1}{(1-\bar{r}^2)^2} d\bar{r}^2 +   
\frac{e^{-w}}{1-\bar{r}^2} \gamma_1. 
\ee
Finally if $C=B=0$ then then $\xi$ is always space--like and does not lead
to an optical structure. Therefore we have
\begin{prop}
\label{reproj}
For any timelike HSO Killing vector, $\xi$, of the metric 
{\em(\ref{metric_final})} where the curvature of $\gamma_0$ is positive, 
the optical metric associated to $\xi$ is diffeomorphic to
\begin{equation}
\bar{h} = \frac{1}{(\phi - \bar{r}^2)^2} d\bar{r}^2 + 
\frac{e^{-w}}{\phi - \bar{r}^2} \gamma_1
\label{desitt}
\end{equation}
for some constant $\phi>0$.
\end{prop}
\section{Projective equivalence}
\label{secprojective}
\subsubsection*{Zero curvature}
We claim that
$\bar{h}$ and $h$ given by (\ref{opt_ko1}) and (\ref{opt_ko2}) respectively
are not  projectively equivalent even up to diffeomorphisms: 
The metric (\ref{opt_ko2}) admits a nontrivial affine equivalence, i. e. 
there exists a covariantly constant symmetric $(0,2)$--tensor  
$h_1$ that is not proportional  to (\ref{opt_ko2}) (in our case $h_1=dr^2$).
The canonical forms of Levi Civita\footnote{The result of Levi--Civita is 
that the metrics 
\[
h=dr^2+f(r)\gamma, \quad\mbox{and}\quad \tilde{h}=\frac{1}{(\kappa f(r)+1)^2}
dr^2 +\frac{f(r)}{\kappa f(r)+1}\gamma
\]
are projectively equivalent for any constant $\kappa$. Here $f$ is an
arbitrary function of $r$ and $\gamma$ is an arbitrary $r$--independent metric.
The result holds in any dimension.}
\cite {LC} implies that (\ref{opt_ko1}) admits a non-affine geodesic equivalence i. e.
there exists  a geodesically equivalent metric   
that is not covariantly constant in the Levi--Civita connection of (\ref{opt_ko1}). 
It is given by
\[
h_2=\frac{1}{\bar{r}^2+1}\Big(\frac{\bar{r}^2}{\bar{r}^2+1}d\bar{r}^2
+e^{-w}\gamma_1\Big).
\]
Thus,
if (\ref{opt_ko1}) and (\ref{opt_ko2}) were equivalent, there
would exist at least three non-proportional metrics sharing the same geodesics.
This in dimension three implies \cite{KM} that $h$ has constant curvature and 
so it is flat\footnote{This example shows that 
some care is needed with the projective Weyl tensor
argument from \cite{GWW08}. Consider the metric  (4.1) in this paper
(numbers as in published version but $r$ replaced by $u$ and $h$ replaced by 
$e^{-w}\gamma_1$)
\[
h=\frac{du^2}{u^4f(u)^2}+\frac{1}{f(u)}e^{-w}\gamma_1.
\]
Taking $f=1$ and  setting $u=1/r$ this gives our (\ref{opt_ko2}). 
Now take (4.1) with $f=2/u$,
so that $u^3f'+(1/2)u^4f''=0$ and the projective Weyl tensor is
the same as that with $f=1$. Changing variables by $u=2/R^2$ gives 
(\ref{opt_ko1}).
So (\ref{opt_ko2}) and (\ref{opt_ko1}) are both of the form (4.1) where 
the Weyl tensor only depends on 
$h_{ij}$ but, as we have demonstrated, 
they are not projective equivalent.}.

\subsubsection*{Non--zero curvature}
Let us first consider the case when $\gamma_0$ has negative curvature.
\begin{prop}
\label{quiv}
Let $\xi_1$ and $\xi_2$ be two timelike HSO Killing vectors for the 
metric $g$ defined by (\ref{metric_final}) where $\gamma_0$ is $AdS_2$. 
Then, the optical metric associated to $\xi_1$ is projectively equivalent 
to the optical metric associated to $\xi_2$ after some diffeomorphism.
Thus all optical metrics are equivalent to (\ref{diffopt}) with $\phi=1$.
\end{prop}
{\bf Proof.} 
Let us first consider 
(\ref{diffopt})
By Proposition \ref{proj} , the optical metric 
associated to any timelike HSO Killing vector $\xi$ is given, after diffeomorphism, by (\ref{diffopt}) for some constant $\phi$. For Killing vectors $\xi_1$ and $\xi_2$, let $h_1$, $h_2$ be the associated optical metrics written in the form (\ref{diffopt}) with corresponding constants $\phi_1$ and $\phi_2$, respectively. Let $\Gamma^{i}_{jk}$, $\tilde{\Gamma}^{i}_{jk}$ be the connection components of the metric connection of $h_1$, $h_2$ respectively. Then, these metrics are projectively equivalent (see, for example, \cite{EM,bryant}) if and only if there exists a one-form $\omega = \omega_j dx^j$ such that
\[
\tilde{\Gamma}^i_{jk} = \Gamma^{i}_{jk} + \delta^{i}_j \omega_k + \delta^i_k \omega_j.
\]
Working this out explicitly, we find that the one-form
\[
\omega =\frac{\bar{r}(\phi_2-\phi_1)}{(\bar{r}^2+\phi_1)(\bar{r}^2+\phi_2)} d\bar{r}
\]
satisfies this criteria.
\koniec
The same argument, with 
\[
\omega =\frac{\bar{r}(\phi_1-\phi_2)}{(\bar{r}^2-\phi_1)(\bar{r}^2-\phi_2)} d\bar{r},
\]
can be used in the $dS_2$ case, to show that any two optical metrics 
(\ref{desitt}) are projectively equivalent to (\ref{desitt}) with $\phi=-1$.
\section{$C$--metric}
\label{seccmetric}
The $C$--metric represents a pair of separated black holes accelerating 
in opposite directions. The original solution  constructed by Weyl
can be generalised to the cosmological setting - the relevant line element with $\Lambda<0$ belongs to the Pleba\'nski--Demia\'nski class
{\cite{Pleb_Dem} and is given by
\be
\label{C_metric}
g=\frac{1}{A^2(x^2+y^2)}\Big(-Fdt^2+\frac{1}{F}dy^2+\frac{1}{G}dx^2
+Gd\phi^2\Big),
\ee
where
\[
F=y^2-2mAy^3+e^2A^2y^4-1-\frac{\Lambda}{3A^2},
\quad
G=1-x^2-2mAx^3-e^2A^2x^4.
\]
The angular coordinate $\phi$ ranges between $-\pi C$ and $\pi C$, where 
$C$ is a positive constant. The constants $A, m$ and $e$ characterise
the acceleration, mass and charge respectively, and are such that
$e^2<m^2$.  The $x$--coordinate lies in an interval between 
two roots of $G$ which contains $0$  and $y\in(-x, \infty)$.

The $C$--metric solves the Einstein--Maxwell equations with the electro--magnetic field $e dy\wedge dt$,
or the pure Einstein equations in the limiting case $e=0$. The case
$m=e=0$ is the space of constant curvature. If $m\neq 0$, the case $A<\sqrt{-\Lambda/3}$ 
corresponds to a single accelerated black--hole and $A>\sqrt{-\Lambda/3}$ 
corresponds to infinite number of pairs of accelerating AdS black--holes.

The optical metric of (\ref{C_metric}) is
\be
\label{opticalC}
h=\frac{1}{F^2}dy^2+\frac{1}{GF}dx^2+\frac{G}{F} d\phi^2.
\ee
We claim that the optical metrics corresponding to different values of $
\Lambda$ are projectively equivalent. To establish this it is enough calculate
the Christoffel symbols of $h$ and notice
that
\[
\Gamma^i_{jk}=(\Gamma_0)^{i}_{jk}+{\delta^i}_j\omega_k+
{\delta^i}_k\omega_j
\]
where $(\Gamma_0)^{i}_{jk}$ is the Levi--Civita connection of (\ref{opticalC}) 
with  $\Lambda=0$ and 
\[
\omega=\omega_i dx^i=\frac{1}{2}d(\ln{(F(y)|_{\Lambda=0})}-\ln{F(y)}).
\]
This projective equivalence implies that the unparametrised geodesics
of $h$ (and so null geodesics of the $C$--metric) are not affected by the 
cosmological constant. The details of this projective equivalence do 
not depend on the exact form of $F=F(y)$ and $G=G(x)$ and the argument above
demonstrates that the projective class does not change under 
$F\rightarrow F+\const$.  Moreover, analysing  
the associated Liouville system \cite{EM,bryant,CD10} it can be shown
any metric which shares unparametrised geodesics with
the optical metric (\ref{opticalC}) is a constant rescaling 
(\ref{opticalC}) possibly with a different value of $\Lambda$.
Setting 
\[
x=\cos{\theta}, \quad y=\frac{1}{Ar}
\]
and taking the limit $A\rightarrow 0$ (need to rescale $t$)
yields the Schwarzschild-de--Sitter metric, and in this case
we recover a known result \cite{Islam,GWW08} that the trajectories of light
rays in the Schwarzchild-de--Sitter metric depend on the mass
but not on the cosmological constant.
\section{Conclusions}
The significance of projective differential geometry in General Relativity 
goes back at least to Weyl: an equivalence class of unparametrised geodesics
can be used to describe the geometry of free falling massive particles. 
Various aspects of the theory have been explored -  
see \cite{HL} and \cite{Matveev} and references therein - but, as emphasised
in \cite{GW10}, there is more to GR than projective geometry. Some cosmological observables - for example cosmic jerk, and its higher order generalisations
\cite{DG} - are not projectively invariant, and thus depend on a choice
of the metric in a projective equivalence class.

 In this paper we have explored a novel aspect of projective equivalence.
The light--rays in static space--times give rise to projective structures
of optical metrics. This leads to ambiguity if a space--time is static
in more than one way, as non--proportional time--like Killing vectors lead to different optical metrics, which as we have demonstrated are not always projectively
equivalent.

\section*{Appendix--ultra static metrics}
\setcounter{equation}{0}
Here we shall show that in the ultra--static case $V=1$, we can integrate the Killing equations
without making the additional genericity assumption 
and establish Proposition (\ref{HSOone}) with $w=\const$ and $\gamma_0$ being flat.
This is essentially the case considered by Sonego
\cite{S10}. We shall however take our analysis further and consider 
optical metrics resulting from this construction. In the adapted coordinate
system, the Killing vector $\tilde{\xi}$ on $\Sigma$ satisfies
\begin{equation}
(\xi^1, \xi^2, \xi^3)\mid_{t=0} = (1, 0, 0).
\label{restriction}
\end{equation}
Now consider the Killing equations for $\xi$. 
Using  $\Gamma_{ij}^0=0$ we find that
$\nabla_{(0} \xi_{0)} = 0, \nabla_{(0} \xi_{i)} = 0 $
imply
\[
\partial_t \xi^0 = 0,\quad
e^{w(x,y)} \partial_t \xi^1 = \partial_r \xi^0,\quad
e^{u(x,y)} \partial_t \xi^2 = \partial_x \xi^0,\quad
e^{u(x,y)} \partial_t \xi^3 = \partial_y \xi^0.
\]
Integrating and using the 
initial conditions (\ref{restriction})  gives
\be
\xi^1 = e^{-w(x,y)} (\partial_r \xi^0) t + 1,\quad
\xi^2 = e^{-u(x,y)} (\partial_x \xi^0) t,\quad
\xi^3 = e^{-u(x,y)} (\partial_y \xi^0) t. 
\label{simpler}
\ee
Now, let us consider the hypersurface orthogonality condition
$\xi\wedge d\xi=0$. We find
\begin{eqnarray*}
0&=&\xi_{[0} \nabla_1 \xi_{2]}\\ 
&=&  - \xi^0 \left( (\partial_r \partial_x \xi^0) t - (\partial_x \partial_r \xi^0) t - \frac{\partial w}{\partial x} e^w \right) \\
&&+ ((\partial_r \xi^0) t + e^w)(-2\partial_x \xi^0) + (\partial_x \xi^0 )t(2\partial_r \xi^0). 
\end{eqnarray*}
This, together with a similar condition resulting from 
$\xi_{[0} \nabla_1 \xi_{3]} = 0$, implies after some algebra
\begin{equation}
\xi^0 = \alpha(r) e^{\frac{1}{2} w(x,y)}.
\label{alphaeq}
\end{equation}
The rest of the hypersurface orthogonality conditions are then satisfied. The remaining Killing equations will yield conditions on $w(x,y)$ and $u(x,y)$ as well as a condition for $\alpha(r)$ as follows: Equation (\ref{alphaeq})
and $\nabla_{(2} \xi_{3)} =0$ give
\begin{equation}
\frac{\partial^2 w}{\partial x \partial y} + \frac{1}{2} \left( \frac{\partial w}{\partial x} \right) \left( \frac{\partial w}{\partial y} \right) = \frac{1}{2} \left[ \left(\frac{\partial u}{\partial x} \right) \left(\frac{\partial w}{\partial y} \right) + \left(\frac{\partial u}{\partial y} \right) \left(\frac{\partial w}{\partial x} \right) \right].
\label{thirteen}
\end{equation}
Similarly, the Killing conditions
$
\nabla_{(2}\xi_{2)} = 0 = \nabla_{(3} \xi_{3)}
$
give 
\begin{eqnarray}
\frac{\partial^2 w}{\partial x^2} + \frac{1}{2} \left(\frac{\partial w}{\partial x} \right)^2 &=& \frac{1}{2} \left[ \left(\frac{\partial u}{\partial x} \right) \left(\frac{\partial w}{\partial x} \right) - \left(\frac{\partial u}{\partial y} \right) \left(\frac{\partial w}{\partial y} \right) \right], \nonumber \\
\frac{\partial^2 w}{\partial y^2} + \frac{1}{2} \left(\frac{\partial w}{\partial y} \right)^2 &=& \frac{1}{2} \left[ \left(\frac{\partial u}{\partial y} \right) \left(\frac{\partial w}{\partial y} \right) - \left(\frac{\partial u}{\partial x} \right) \left(\frac{\partial w}{\partial x} \right) \right].
\label{fourteen}
\end{eqnarray}
The Killing equations
$
\nabla_{(1} \xi_{2)} = 0 = \nabla_{(1} \xi_{3)}
$
are now satisfied and the condition
$
\nabla_{(1} \xi_{1)} = 0
$
gives 
\[
\frac{\partial^2 \alpha(r)}{\partial r^2} = -\frac{1}{4} e^{w-u} \left[ \left(\frac{\partial w}{\partial x} \right)^2 + \left( \frac{\partial w}{\partial y} \right)^2 \right] \alpha(r).
\]
The left hand side of this equation depends only on $r$, so the quantity
\begin{equation}
\mu^2 \equiv \frac{1}{4} e^{w-u} \left[ \left( \frac{\partial w}{\partial x} \right)^2 + \left( \frac{\partial w}{\partial y} \right)^2 \right]
\label{omeger}
\end{equation}
is a constant. Let us first consider the case ${\mu \neq 0}$. Solving (\ref{omeger}) for $u$ and substituting the partial derivatives
of $u$ into (\ref{thirteen}) and (\ref{fourteen}) gives, after some algebra,
\[
\frac{\partial^2 w}{\partial x^2}+ \frac{\partial^2 w}{\partial y^2}+ \frac{1}{2}\left(\left(\frac{\partial w}{\partial x}\right)^2 + \left(\frac{\partial w}{\partial y} \right)^2 \right) = 0.
\]
This means that the function $e^{w/2}$ is  harmonic,  thus
$
e^{w(x, y)/2} = G(z) +  \overline{G(z)},
$
where $G$ is holomorphic in $z=x+iy$. A coordinate transformation
\[
X=\frac{2}{\mu}\mbox{Re}\;(G)\;\cos{(\mu r)}, \quad Y=\frac{2}{\mu}\mbox{Re}\;(G)\;\cos{(\mu r)}, \quad Z=\frac{2}{\mu}\mbox{Im}\;(G), \quad T=t
\]
yields the Minkowski metric $g=-dT^2+dX^2+dY^2+dZ^2$.
\vskip5pt
 Now, let us consider the case ${\mu=0}$.
Equation (\ref{omeger}) implies that $w(x,y)$ is a constant and so 
the metric (\ref{canonical}),
after rescaling $r$, becomes
\[
g = -dt^2 + dr^2 + \gamma_1,
\]
where $\gamma_1=e^{u}(dx^2+dy^2)$. We also have $\alpha=Ar+B$, and
given the initial conditions, the Killing vector $\xi$ can be written as
\[
\xi = (Ar + B e^{\frac{1}{2} w}) \frac{\partial}{\partial t} + (A t + e^{\frac{1}{2} w} ) \frac{\partial}{\partial r}.
\]
If $A\neq 0$ we translate $(r, t)$ by adding constants and rescale the Killing vector so that 
\[
\xi=r\frac{\p}{\p t}+t\frac{\p}{\p r}.
\]
Setting $t=\bar{r}\sinh{(\bar{t})}, r=\bar{r}\cosh{(\bar{t})}$ gives 
\[
g=\bar{r}^2(-d\bar{t}^2+\bar{h})
\]
where 
\be
\label{lab5}
\bar{h}=\bar{r}^{-2}(d\bar{r}^2+\gamma_1)
\ee
is the optical metric associated to the Killing vector $\p/\p \bar{t}$.
\vskip5pt
If $A=0$ then a constant rescaling of $t$ can be used to set 
$\xi=\cos{\theta}\p_t+\sin{\theta}\p_r$, where $\theta$ is a constant
in a range which makes  $\xi$ is time--like. The pseudo--orthogonal transformation 
of $(r, t)$ can now be used to set $\xi=\p/\p t$, so the optical metric in this case is
\be
\label{opt_1_us}
h=dr^2+\gamma_1.
\ee


\begin{thebibliography}{jafsdl}
\frenchspacing 

\bibitem{abramowicz} Abramowicz, M. A., Carter, B., \and Lasota, J. P. (1988)
Optical Reference Geometry for Stationary and Static Dynamics.
GRG {\bf 20}, 1173.

\bibitem{bryant} Bryant, R. L., Dunajski, M. \and Eastwood, M. (2009) 
Metrisability of two-dimensional projective structures. 
J. Differential Geometry. {\bf 83}, 465-499. 


\bibitem{CD10} Casey, S. \and Dunajski, M. (2011) 
Metrisability of Path Geometries. In preparation.

\bibitem{DG} Dunajski, M. \and Gibbons, G. (2008)
Cosmic jerk, snap and beyond. 
Class. Quantum Grav.  {\bf 25}, 235012.

\bibitem{EM} Eastwood, M. G. \&  Matveev, V (2007)
Metric connections in projective differential geometry, in
\emph{Symmetries and Overdetermined Systems of Partial Differential Equations},
IMA Volumes in Mathematics and its Applications 144, 
Springer Verlag 2007, pp.~339--350

\bibitem{GWW08}
Gibbons, G. W., Warnick C. M. \and Werner M. C. (2008)
Light-bending in Schwarzschild-de-Sitter: 
projective geometry of the optical metric.
Class. Quant. Grav. {\bf 25}, 245009.

\bibitem{GW}  Gibbons, G. W. \and Warnick C. M. (2009) 
Universal properties of near-horizon optical geometry. 
Phys.Rev.{\bf D79}:06403.

\bibitem{GW10} Gibbons, G. W.; Warnick, C. M. (2010) Dark energy 
and projective symmetry. Phys. Lett. {\bf B 688}, 337-340.

\bibitem{HE} Hawking, S.W. \and Ellis, G. F. R. (1973) 
{\em The Large Scale Structure of Space-Time}, CUP, Cambridge.

\bibitem{HL} Hall, G. S. \and Lonie, D. P (2008) 
The principle of equivalence and cosmological metrics, 
J. Math. Phys. {\bf 49} 022502.


\bibitem{Islam} Islam, J. N. (1983) The cosmological constant and classical tests of general relativity.


\bibitem{KM} Koisak, V. \and Matveev. V. S. (2008)
Complete Einstein metrics are geodesically rigid.
{\tt arXiv:0806.3169}.

\bibitem{LC} Levi--Civita, T. (1896) On the transformations of the 
dynamical equations. Annali di Mathematica, {\bf 24}, 255-300.

\bibitem{Matveev}  Matveev. V. S. (2011)
Geodesically equivalent metrics in general relativity.
{\tt  arXiv: 1101.2069}.

\bibitem{Pleb_Dem} Pleba\'nski, F. \and Demia\'nski, M. (1976)
Rotating, charged, and uniformly accelerating mass in general relativity.
Annals Phys. {\bf 98}: 98-127.

\bibitem{S10} Sonego, S. (2010)
Ultrastatic spacetimes.  J. Math. Phys. {\bf 51}, 092502. 


\end{thebibliography}
\end{document}